\documentclass[journal,utf8]{IEEEtran}
\usepackage[utf8]{inputenc}
\usepackage[pdftex]{graphicx}
\usepackage[numbers,sort&compress,square]{natbib}
\usepackage{amsmath}
\usepackage[caption=false,font=footnotesize]{subfig}
\usepackage{url}

\begin{document}

\bstctlcite{MyBSTcontrol}

\title{Urban scaling of football followership on Twitter}

\author{Eszter~Bok\'anyi,~Attila~S\'oti~and~G\'abor~Vattay%
\thanks{E. Bok\'anyi and G. Vattay are at the Department of Physics of Complex Systems, E\"otv\"os Lor\'and University, Budapest Hungary.}%
\thanks{A. S\'oti is the member in the Doctoral School of Regional Sciences and Business Administration at Sz\'echenyi Istv\'an University, Gy\"or, Hungary.}%
}

\markboth{Journal of \LaTeX\ Class Files,~Vol.~14, No.~8, August~2015}%
{Shell \MakeLowercase{\textit{et al.}}: Bare Demo of IEEEtran.cls for IEEE Journals}

\maketitle

\begin{abstract}
Social sciences have an important challenge today to take advantage of new research opportunities provided by large amounts of data generated by online social networks. Because of its marketing value, sports clubs are also motivated in creating and maintaining a stable audience in social media. In this paper, we analyze followers of prominent footballs clubs on Twitter by obtaining their home locations. We then measure how city size is connected to the number of followers using the theory of urban scaling. The results show that the scaling exponents of club followers depend on the income of a country. These findings could be used to understand the structure and potential growth areas of global football audiences.
\end{abstract}

\begin{IEEEkeywords}
urban scaling, Twitter, social media, football
\end{IEEEkeywords}

\section{Introduction}
\IEEEPARstart{T}{oday} the online social network Twitter has more than 300 million monthly active users \cite{Twitter2018}, with many of them actively following sports events, stars or clubs to exploit the possibilities of obtaining the latest news through instantaneous messaging \cite{Bruns2014}. Large football clubs and football leagues invest money in establishing official social media channels to engage with their fan basis \cite{Price2013}, and seek to purchase players who bring them a massive number of Twitter followers \cite{KpmgRonaldo}. Social media presence is especially important for clubs that rely more heavily on broadcasting and commercial revenues than on match day revenues, such as the global top 20 clubs from a recent analysis of the Deloitte Football Money League \cite{Deloitte2018}. Because global fans have limited options to be present at matchday events, popularity on Facebook together with Twitter is a good indicator to judge the global follower success of a football club.

On the other hand, the geographic and socio-economical environment of a user still plays an important role in determining the probability of engaging with a globalized phenomenon. As such, complex spatial structures and the dynamics of changes in them have for some time been a focus of the scientific community as well as marketing experts. Recently, there has been a growing literature on the concept of urban scaling, which connects measurable outputs of cities to their size \cite{Bettencourt2010d,Alves2015c, Arcaute2015a,Cottineau2017,Cottineau2018DefiningEconomies,Bettencourt2013a,Bettencourt2013d,Gomez-Lievano2012,Yakubo2014SuperlinearCities}. Urban scaling laws have been detected for various quantities with respect to city size, such as GDP \cite{Lobo2013}, urban economic diversification \cite{Strumsky2016}, touristic attractiveness \cite{Bojic2016ScalingStates}, crime concentration \cite{Oliveira2017,Hanley2016a}, human interactions \cite{Schlapfer2014b}, election data \cite{Bokanyi2018UniversalResults} or even building heights \cite{Schlapfer2015UrbanSize}. Some of these measures follow a superlinear relationship with urban size, which means that the quantities are disproportionally overrepresented in larger cities. These measures include GDP, number of patents or certain business types, where larger cities facilitate more the accumulation of wealth and resources needed for such phenomena. On the other hand, infrastructural-like quantities have sublinear scaling laws reflecting efficiency due to urban agglomeration effects.

In this paper, we investigate urban scaling laws for geolocated Twitter football club followers for three major widely acknowledged clubs: Real Madrid, Manchester United and Bayern Munich. We calculate the scaling exponents for the number of followers of each club in the urban systems of five different countries. While the scaling exponents of clubs differ significantly within countries as well, the variations in the exponents across countries suggest that the more wealthier a country is, the more sublinear its follower scaling exponent, and vice versa.

\section{Materials and methods}

Twitter freely provides approximately 1\% of its data for download through its API. For those users that allow this option on their smartphones, the exact GPS coordinates are attached to their messages, the so-called tweets. By focusing the data collection on these geolocated tweets, we could determine the home location for a selected most active users in the database using the friend-of-friend algorithm clustering on their coordinated messages. This left us with a total of 26.3 million Twitter users that have home coordinates associated to them. We constructed a geographically indexed database of these users, permitting the efficient analysis of regional features \cite{Dobos2013}. Using the Hierarchical Triangular Mesh scheme for practical geographic indexing, we assigned cities to each user. City locations were obtained from \url{http://geonames.org}, city bounding boxes via the Google Places API.

We downloaded the Twitter user identifiers of the followers of three selected football clubs: Real Madrid, Manchester United and Bayern Munich. Table 1 shows the number of followers (people who follow at least one of the three teams, later referred as overall follower count) that are also in our geolocated user database, which meant roughly 2-3\% of all followers in all three cases.

\begin{table}[!ht]
  \begin{center}
    \label{tab:table1}
    \begin{tabular}{l|c|r}
      \textbf{Team name} & \textbf{Total number of followers} & \textbf{Geolocated followers}\\
      \hline
      Real Madrid & 28.7M & 808,427\\
      Manchester United & 17.3M & 436,515\\
      Bayern Munich & 4.3M & 119,056\\
    \end{tabular}
  \end{center}
  \caption{Number of total followers for each football club on Twitter and number of followers from the geolocated user database used in our analysis.}
\end{table}

The theory of urban scaling \cite{Alves2015c} suggests that there is a power-law relationship between a socio-economic indicator measured in a city and its size. We can formulate this power-law relationship with the following equation:
 \begin{equation}
Y=Y_0N^\beta,
\end{equation}

Where $Y$ denotes the investigated quantity, $N$ is the number of inhabitants in a city, $Y_0$ is a normalization constant, and $\beta$ is the so-called exponent that characterizes the behavior of the quantity in connection to changing city size. In the literature, it has been observed, that this $\beta$ parameter differs only slightly from 1. Most urban socio-economic indicators have superlinear $\beta>1$ exponents, which is caused by larger cities being the centers of wealth, innovation and creative processes. Sublinear scaling $\beta<1$ characterizes material quantities associated with infrastructure, where the agglomeration into cities is more economic, which manifests in fewer overall road length, or overall cable need etc. \cite{Bettencourt2010d}.

If we take the logarithm of both sides, the equation becomes a linear relationship:
\begin{equation}
\log Y = \log Y_0 + \beta \cdot \log N, 
\end{equation}

It is then enough to fit a line onto the $\log Y$ -- $\log N$ pairs. We used binning of the data, where we took the mean of  $log N$ and $log Y$ in each bin, and then fitted a line onto them using an OLS fit with weighting the bins by $1/\sqrt{N}$. This error calculation assumes that higher follower numbers carry less error when fitting the scaling curves \cite{Bettencourt2007}.

\begin{figure}[!t]
\includegraphics[width=3.5in]{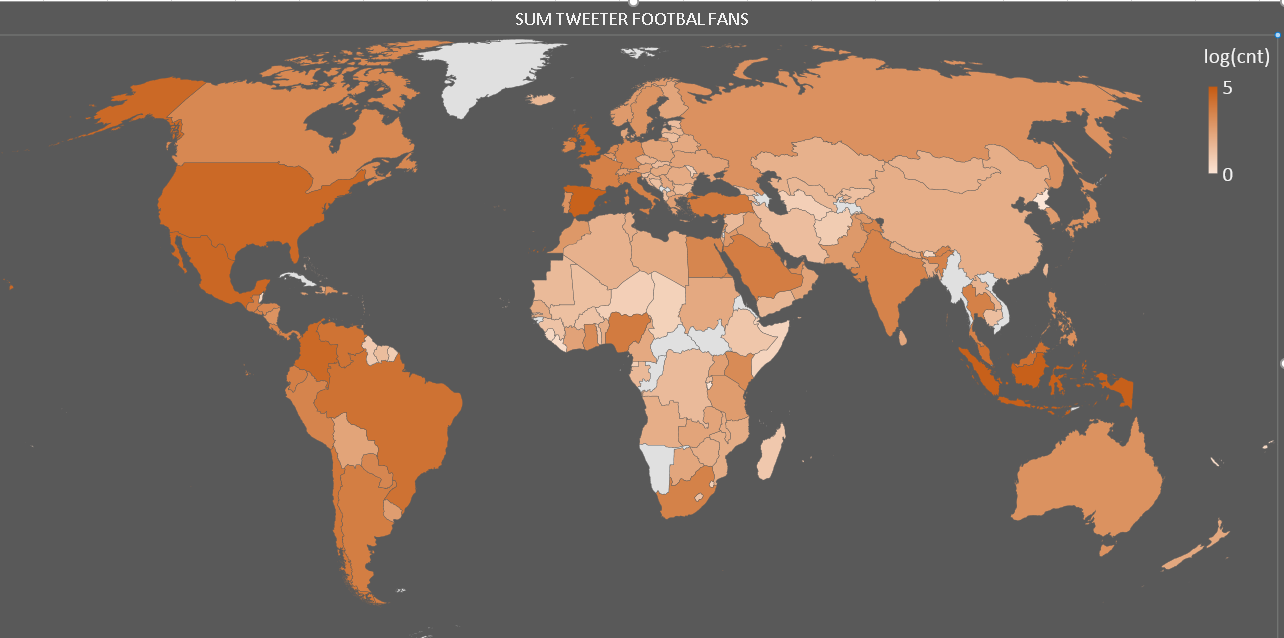}
\caption{\bf {Distribution of geolocated Twitter users that follow at least one of the three selected clubs. Countries are colored according to the logarithm of the number of users.}}
\label{fig3}
\end{figure}

\section{Results and discussion}

The geographical distribution of users that follow at least one of the three clubs can be seen in Figure~\ref{fig3}. A major fan base is in Western Europe, North and Latin America as well as in the Pacific Region. Because Spanish and English teams are among the investigated clubs in Spain and in Great Britain the number of followers is high. 

As analyzed countries, we chose the home countries of two of the teams, Spain and the UK, and we included traditional football supporter countries like Mexico. We chose Indonesia from from the Pacific Region, and Columbia from South America. We also analyze the USA since it is a country with high Twitter penetration.

\begin{figure}
\centering
\includegraphics[width=3.5in]{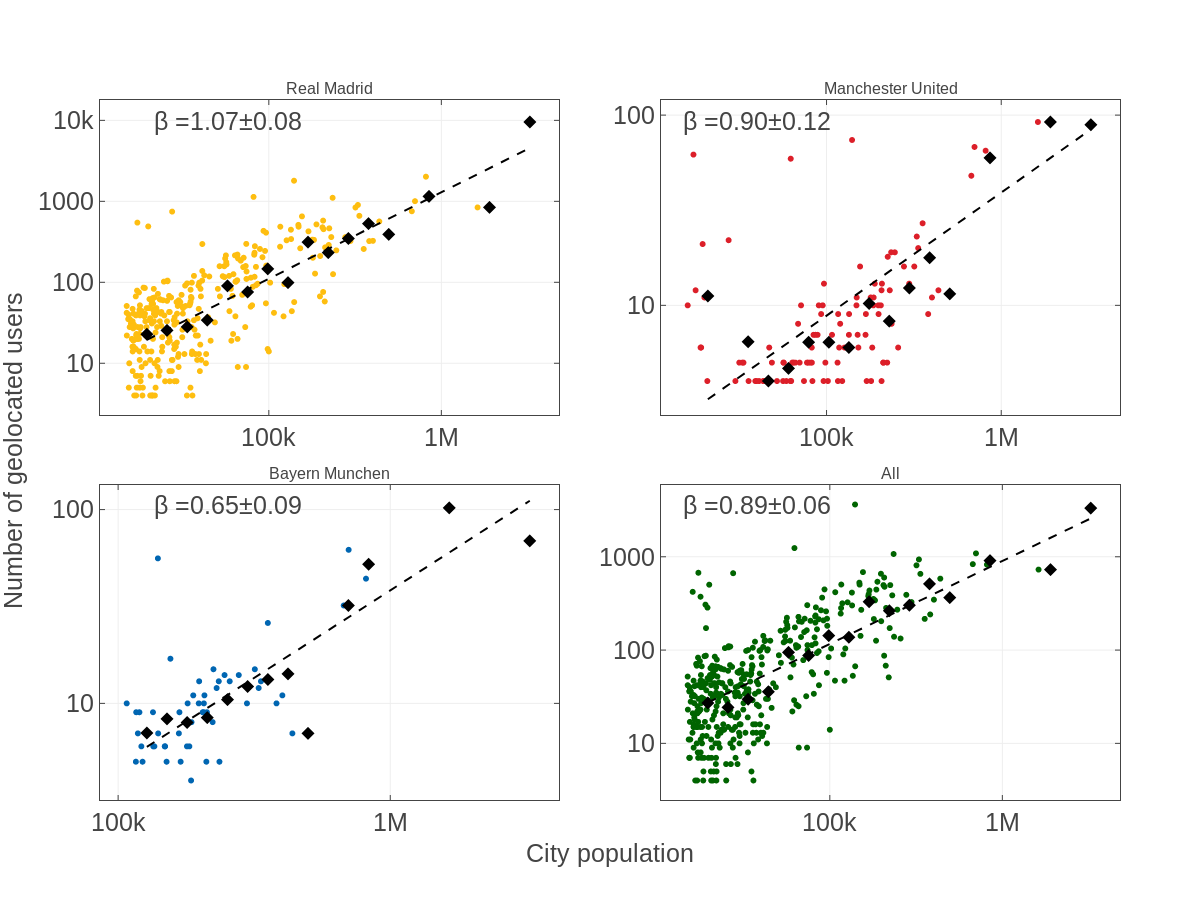}
\caption{ \bf Number of followers for the three selected clubs (A-C), and combined follower number (D) as a function of city size in Spain. Black diamonds correspond to bin averages, dashed lines represent the OLS fits with exponents $\beta_{RM}=1.07\pm0.08$, $\beta_{MU}=0.90\pm0.12$, $\beta_{BM}=0.65\pm0.09$ and $\beta_{All}=0.89\pm0.06$, respectively.}
\label{fig4}
\end{figure}

In the top left corner of Figure~\ref{fig4}, we can see the urban scaling relationships of Spain for the three clubs (Real Madrid in the top left, Manchester United in the top right and Bayern Munich in the bottom left corner), and for the number of overall followers (bottom right corner). The exponent of Real Madrid, the "home" team is superlinear ($\beta_{RM}=1.07\pm0.08$), while the exponent of the other two teams are sublinear with $\beta_{MU}=0.90\pm0.12$ for the Manchester United, and $\beta_{BM}=0.65\pm0.09$ for the Bayern Munich, respectively. Is is spectacular how the second biggest city in Spain, Barcelona is a clear outlier in the Real Madrid urban scaling curve, with having much less followers than the size of the city would predict. The overall follower numbers in Spain also have a sublinear scaling.

\begin{figure}
\centering
\includegraphics[width=3.5in]{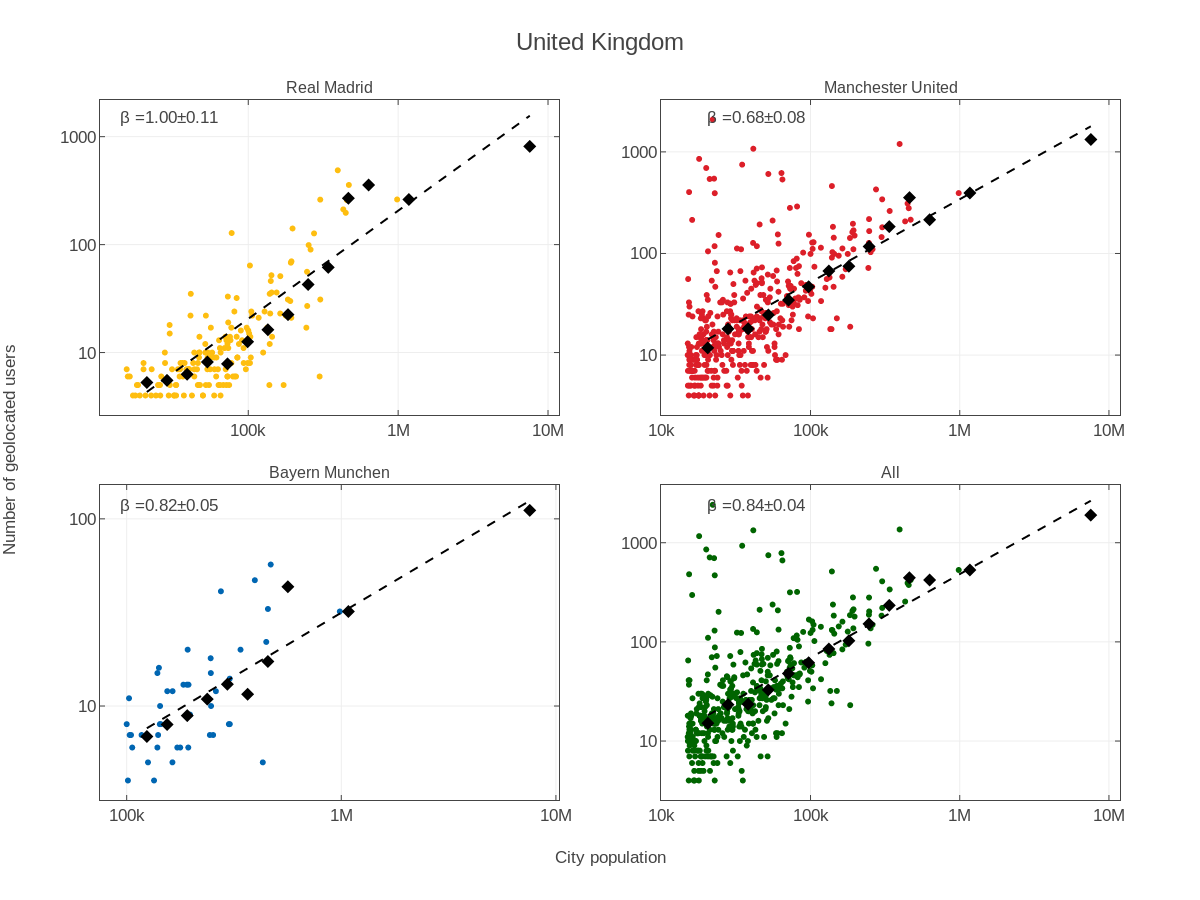}
\caption{\bf Number of followers for the three selected clubs (A-C), and combined follower number (D) as a function of city size in the UK. Black diamonds correspond to bin averages, dashed lines represent the OLS fits with exponents $\beta_{RM}=1.00\pm0.11$, $\beta_{MU}=0.68\pm0.08$, $\beta_{BM}=0.82\pm0.05$ and $\beta_{All}=0.84\pm0.04$, respectively.}
\label{fig5}
\end{figure}

In Figure~\ref{fig5} when we look at scaling curves in the UK, which has the longest football traditions of all of the countries, we again see a similar picture of the exponents, with that of Real Madrid being higher than the other two, though it is only around the linear regime with $\beta_{RM}=1.00\pm 0.11$. However, Manchester United, apart from the outlier points of Manchester and its surroundings has an astoundingly low sublinear exponent $beta_{MU}=0.68\pm0.08$ that suggests a strong relative decline of interest for this team with the city size. The overall follower trend is also strongly sublinear in the UK.

\begin{figure}
\centering
\includegraphics[width=3.5in]{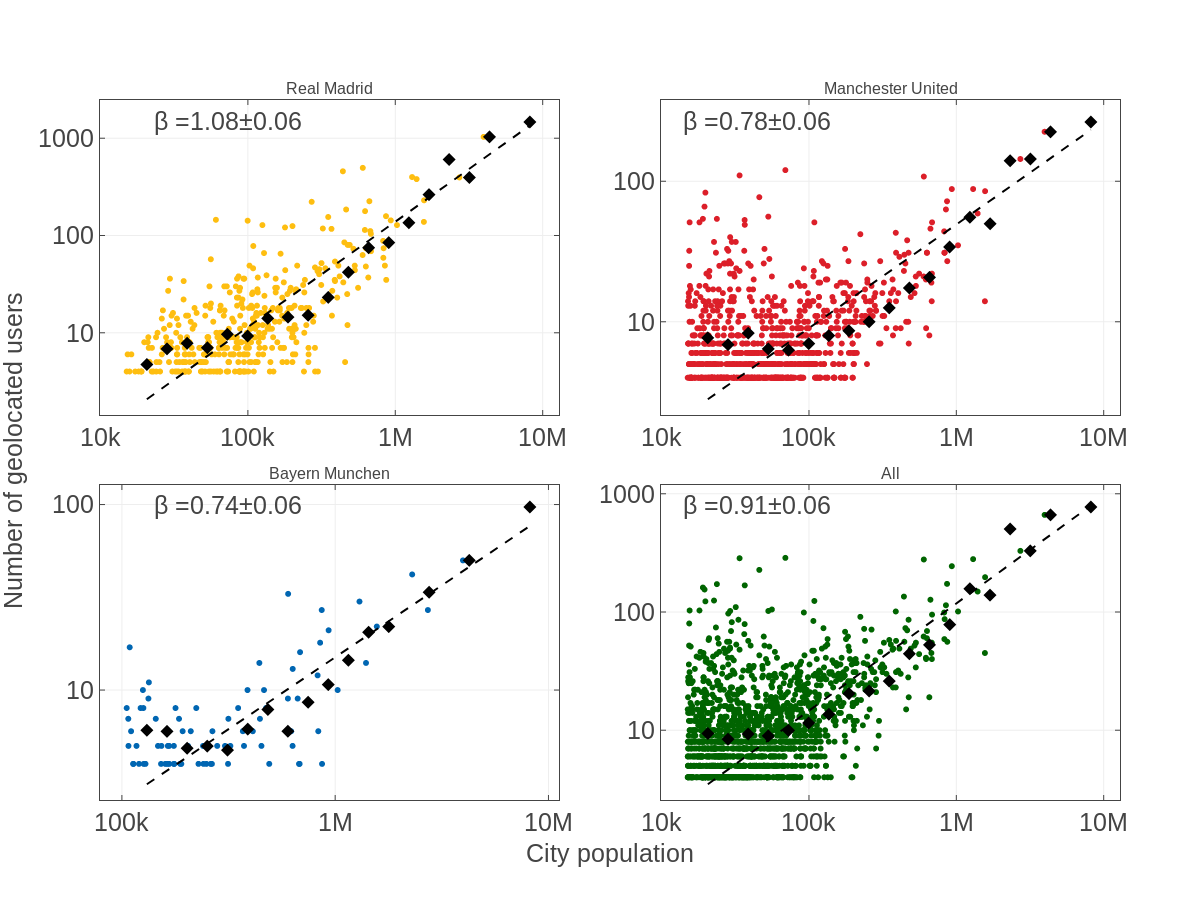}
\caption{\bf Number of followers for the three selected clubs (A-C), and combined follower number (D) as a function of city size in the US. Black diamonds correspond to bin averages, dashed lines represent the OLS fits with exponents $\beta_{RM}=1.08\pm0.06$, $\beta_{MU}=0.78\pm0.06$, $\beta_{BM}=0.74\pm0.06$ and $\beta_{All}=0.91\pm0.06$, respectively.}
\label{fig6}
\end{figure}

The case of the USA in Figure~\ref{fig6} is very similar to that of Spain, where Real Madrid followers scale superlinearly, but the other two clubs have a sublinear relationship with city size.

\begin{figure}
\centering
\includegraphics[width=3.5in]{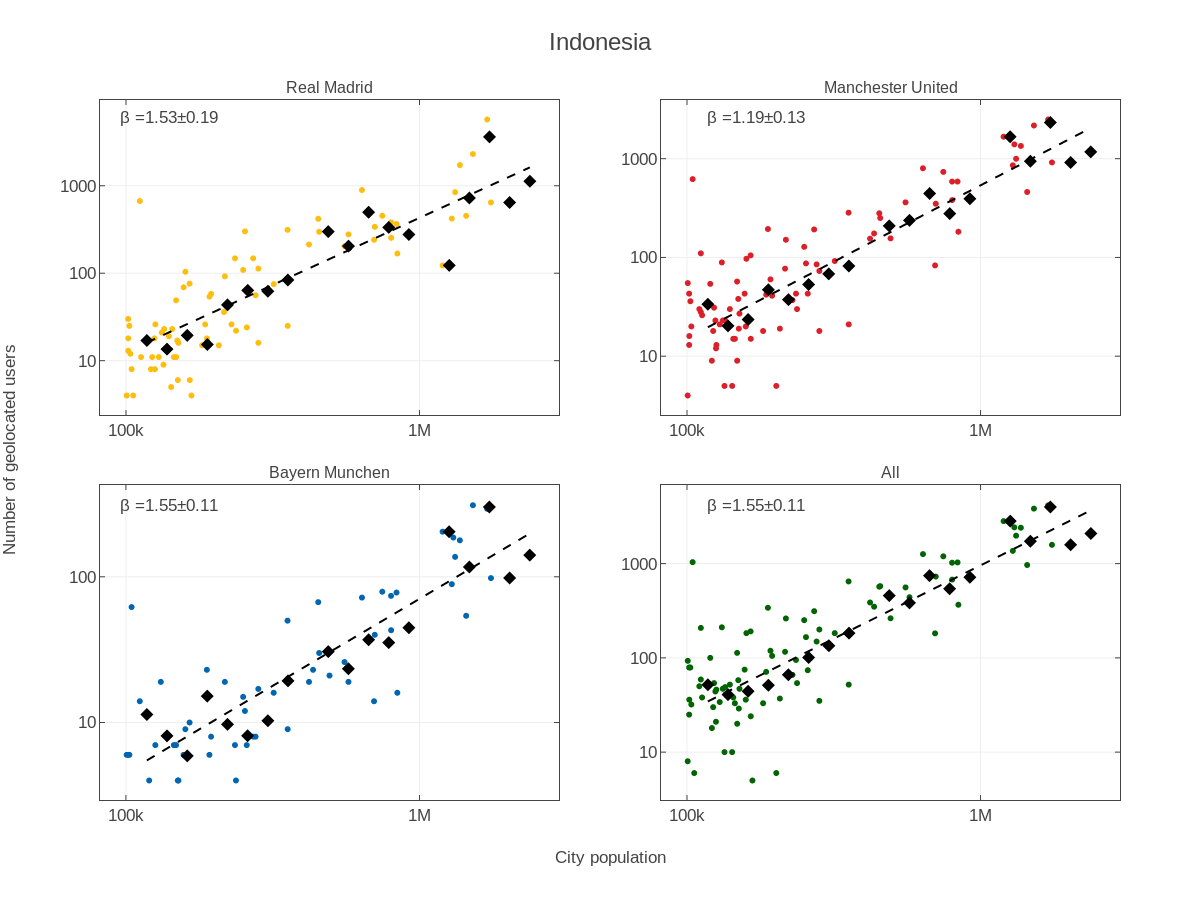}
\caption{ \bf Number of followers for the three selected clubs (A-C), and combined follower number (D) as a function of city size in Indonesia. Black diamonds correspond to bin averages, dashed lines represent the OLS fits with exponents $\beta_{RM}=1.53\pm0.19$, $\beta_{MU}=1.19\pm0.13$, $\beta_{BM}=1.55\pm0.11$ and $\beta_{All}=1.55\pm0.11$, respectively.}
\label{fig7}
\end{figure}

\begin{figure}
\centering
\includegraphics[width=3.5in]{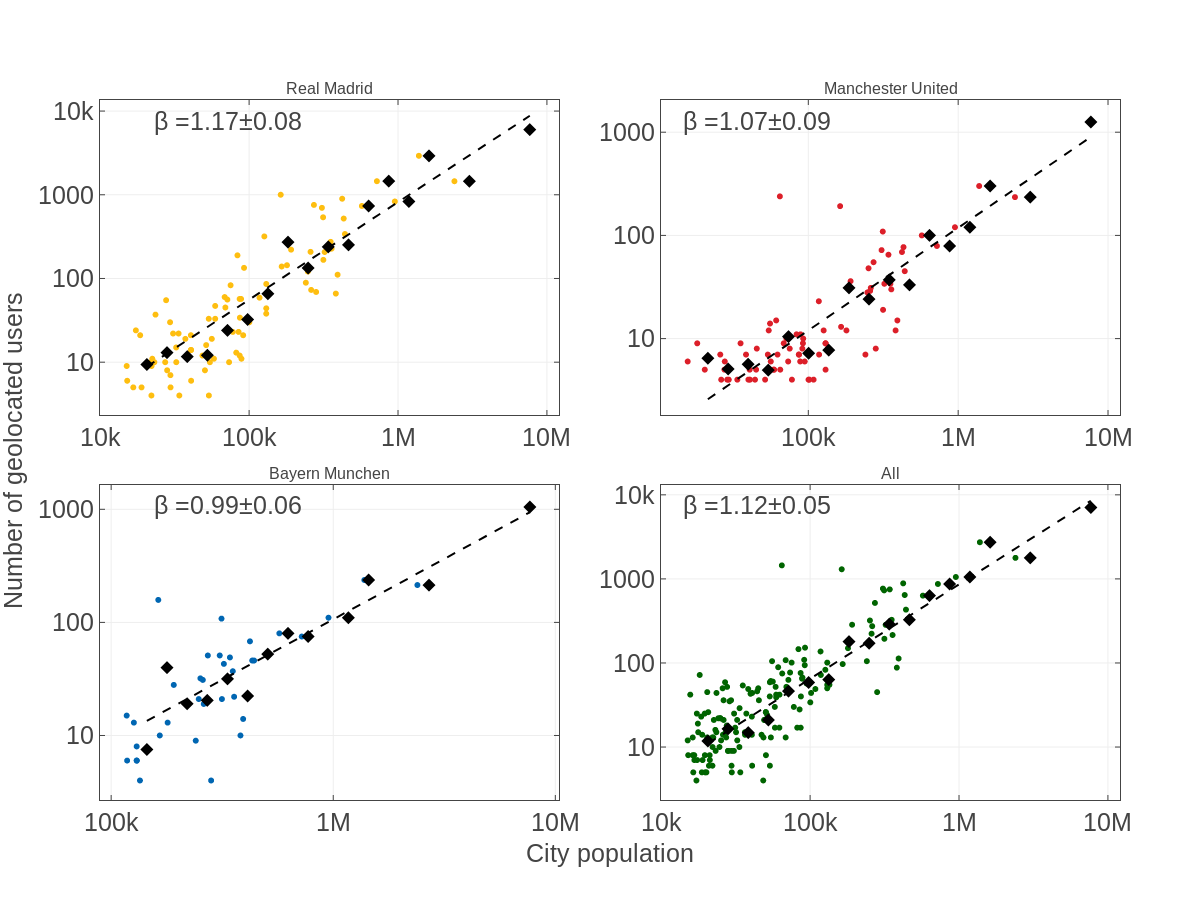}
\caption{\bf Number of followers for the three selected clubs (A-C), and combined follower number (D) as a function of city size in Columbia. Black diamonds correspond to bin averages, dashed lines represent the OLS fits with exponents $\beta_{RM}=1.17\pm0.08$, $\beta_{MU}=1.07\pm0.09$, $\beta_{BM}=0.99\pm0.06$ and $\beta_{All}=1.12\pm0.05$, respectively.}
\label{fig8}
\end{figure}

A very different effect takes place in Indonesia according to Figure~\ref{fig7}. Here, all four scaling relationships are in the highly superlinear range, which means that club followership is a measure that is driven by urban factors. Though less pronounced because of slightly smaller, but still superlinear exponents, this is also the case for Columbia in Figure~\ref{fig8}.

\begin{figure}[!ht]
\centering
\includegraphics[width=3.5in]{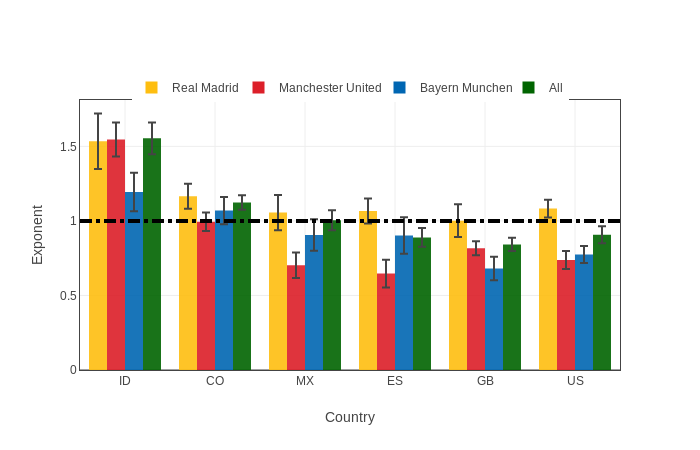}
\caption{ \bf Summary figure showing the exponents per team according to countries. The vertical line at $\beta=1$ corresponds to linear scaling.}
\label{fig9}
\end{figure}

The summary Figure~\ref{fig9} shows that Columbia, Indonesia and Mexico, are the countries whose exponents for the overall supporter count are superlinear. This means that in these countries that globalized football tracking is an increasingly urban phenomenon. In countries where football culture is older, and/or general income is higher, sublinear exponents may signal a relative attention shift for football to smaller settlements, and a change in the composition of consumers of football-related content. This may be an important message for marketers trying to increase social media attention and responsiveness, because people from different environments may need quite different targeting messages.

The difference between the club exponents in the same country suggests that Real Madrid followers are relatively more abundant in bigger cities, and the other two teams have more or less the same exponents. This suggests that even within a country, different clubs may have different follower audiences and fans from different cultural backgrounds.

\section{Conclusion}

In this paper, we analyzed urban scaling in the follower numbers of three football clubs, Real Madrid, Manchester United and Bayern Munich. We determined user geolocation from Twitter messages that had GPS coordinates attached to them, and fitted scaling relationships using population data for cities of six different countries. While for higher-income countries, urban scaling exponents tended to be in the sublinear, linear or in a few cases, a slightly superlinear range, exponents for lower-income countries are almost exclusively superlinear. This suggests that in a globalized football fandom, very different factors drive followership. Exponents also exhibited variations between clubs, which suggests that the followers of different football clubs are embedded into different socio-economical environments that is related to the degree of urbanization as well.

\section*{Acknowledgment}

The authors thank the support of the National Research, Development and Innovation Office of Hungary (grant no. KH125280).

\bibliographystyle{IEEEtran}
\bibliography{biblio}

\end{document}